\documentclass[aps,prl,twocolumn,showpacs,groupedaddress]{revtex4}
\usepackage{dcolumn}
\usepackage{graphicx}
\usepackage{bm}
\begin{document}


\title{Hydration-induced anisotropic spin fluctuations in Na$_{x}$CoO$_{2}\cdot$1.3H$_{2}$O superconductor}

\author{K. Matano$^1$}
\author{C.T. Lin$^2$}
\author{Guo-qing Zheng$^1$}

\affiliation{$^1$Department of Physics, Okayama University, Okayama 700-8530, Japan}
\affiliation{$^2$Max Planck Institute, Heisenbergstrasse 1, D-70569 Stuttgart, Germany}




\date{Feb.6, 2008}

\begin{abstract}
We report $^{59}$Co NMR studies in single crystals of cobalt oxide superconductor Na$_{0.42}$CoO$_{2}\cdot$1.3H$_{2}$O ($T_c=$4.25K) and its parent compound
 Na$_{0.42}$CoO$_{2}$. 
We find that both the magnitude and the temperature ($T$) dependence of the Knight shifts are identical in the two compounds above $T_c$. 
The spin-lattice relaxation rate ($1/T_1$) is also identical above $T_0 \sim$60 K for both compounds. Below $T_0$,   the  unhydrated sample  is found to be a non-correlated metal that well conforms to Fermi liquid theory, while spin fluctuations develop  in the superconductor. These results indicate that  water intercalation  does not change the density of states but its primary role is to bring about spin fluctuations.
Our result  shows that, in the hydrated superconducting compound, the in-plane spin fluctuation around finite wave vector is  much stronger than that along the $c$-axis, which indicates that the spin correlation is quasi-two-dimensional. 

\end{abstract}

\pacs{74.25.Jb, 74.70.-b, 74.25.Nf}

\maketitle


The hydrated cobalt-oxide  superconductor Na$_{x}$CoO$_{2}\cdot$ 1.3H$_{2}$O 
has attracted much attention in the past few years \cite{Takada}.
This compound bears similarities to the high transition-temperature ($T_c$) copper oxides in that it has layered crystal structure and contains transition-metal element that carries a spin of $\frac{1}{2}$. Moreover, Co forms a triangular lattice rather than a square lattice as in the cuprates, which may lead to new physics associated with spin frustrations. Therefore, insights into this new class of superconductors are expected to shed light on the mechanism of high-$T_c$ superconductivity and may also have impact on other strongly correlated electron systems.

Nuclear quadrupole resonance (NQR) has revealed the unconventional nature of the superconductivity; 
 the spin-lattice relaxation rate $1/T_1$ shows no coherence peak just below $T_c$, which is suggestive of non-s-wave superconducting state \cite{Fujimoto}. The absence of the coherence peak is universal irrespective of Na content or $T_c$ value  \cite{Zheng,Kusano}. At low temperatures, $1/T_1$ follows a $T^3$ variation down to a temperature as low as $\frac{T_c}{9}$ \cite{Zheng,Kusano}, which indicates the existence of nodes (zeroes) in the gap function. Knowledge about  spin-pairing symmetry was obtained from Knight shift \cite{Zheng2,Kobayashi}.
Precise measurements of the Knight shift in a high quality single crystal reveals that the spin susceptibility decreases below $T_c$ along both $a$- and $c$-axis directions, which indicates that the Cooper pairs are in the spin-singlet  state \cite{Zheng2}. Thus, the superconductivity appears to be of nodal $d$-wave symmetry\cite{Eremin,ZhouWang}.

In the normal state, the quantity $1/T_1T$ does not follow the Korringa relation for a non-correlated metal, but increases with decreasing temperature, which indicates electron correlations \cite{Fujimoto}. $^{59}$Co Knight shift measurement shows that the uniform spin susceptibility decreases with decreasing temperature and becomes a constant below $T\sim$ 60 K before superconductivity sets in, along both $a$- and $c$-axis directions \cite{Zheng2}. This result strongly suggests that the spin correlations are of  antiferromagnetic (AF) origin. 
It is worth noting that such AF-like spin fluctuations   increase with decreasing Na-content,
and become strongest at $x \sim$ 0.26 where $T_c$ is the highest \cite{Zheng}.

In cuprates,  isotropic spin fluctuations  have been revealed  by neutron scattering measurements  \cite{Kastner}, which provided a basis for exploring  the relationship between electron correlation and the occurrence of superconductivity. However, the detailed characters of the spin fluctuations, for example, its anisotropy, is unknown in the cobaltates. This is due mainly to the difficulty in growing large size single crystals. The other outstanding issue in the cobaltate is  the role of water intercalation  in achieving superconductivity. In spite of  efforts \cite{Kotliar,Johannes,Takada2,Milne,Muk}, it remains unclear. Also, researches on unhydrated Na$_x$CoO$_2$ compounds have so far focused on high Na-concentrations with  $x\geq$0.5; the electronic state of Na$_x$CoO$_2$ ($x<$0.5), the parent of the superconductor, is poorly understood. Previous NQR or NMR measurements did not provide detailed information on the electron correlations \cite{Ning, IshidaK}.

In this Letter, we address these issues through  comparative studies of the Knight shift and $1/T_1$ on 
 hydrated superconductor Na$_{0.42}$CoO$_{2}\cdot$1.3H$_{2}$O and unhydrated Na$_{0.42}$CoO$_{2}$ single crystals. We find that Na$_{0.42}$CoO$_{2}$ is a conventional metal that well conforms to  Fermi liquid theory for non-correlated electrons, and that hydration  does not alter the density of states (DOS) or carrier concentration. The principal role of hydration is to  bring about  spin fluctuations which are found to be highly anisotropic.  The quasi-2D spin correlations appear to be essential for the superconductivity.

The single crystals of Na$_{0.42}$CoO$_{2}$  and Na$_{x}$CoO$_{2}\cdot$ 1.3H$_{2}$O used in this study were grown by the traveling solvent floating zone (TSFZ) method, as described in  previous publications \cite{Chen,Lin}.
The $^{59}$Co NMR spectra were taken by changing the external magnetic field ($H$) at a fixed rf frequency and recording the echo intensity step by step.
The $1/T_1$  was measured at the lower first-satellite ($3/2\leftrightarrow 1/2$ transition) for $H \parallel c$, and at the central peak ($1/2\leftrightarrow -1/2$ transition) for $H \parallel a$.
The value of  $1/T_1$ was extracted by fitting the nuclear magnetization to the theoretical curve of  Narath \cite{Narath}.

\begin{figure}[h]
\includegraphics[width=7cm]{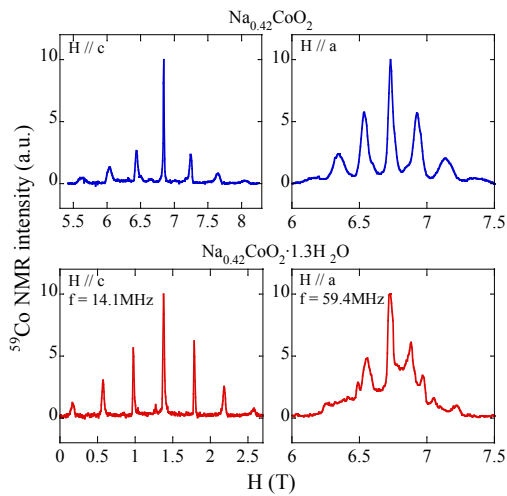}
\caption{\label{fig:spec} 
$^{59}$Co NMR spectra in Na$_{0.42}$CoO$_{2}$  (upper panel)
and Na$_{0.42}$CoO$_{2}\cdot$ 1.3H$_{2}$O (lower panel) taken at   $T$=4.2 K. The NMR frequency is 14.1 MHz and 70.1 MHz for $H \parallel c$  and $H \parallel a$, respectively. The magnetic field was calibrated by $^{27}$Al NMR of  aluminium metal. }
\end{figure}

Figure 1 shows a typical example of the NMR spectra for both samples with the magnetic field applied along 
 the $c$- ($H \parallel c$) and $a$-axis ($H \parallel a$).
For $H \parallel c$, a sharp central transition line accompanied  by six satellites due to the nuclear quadrupole interaction is observed.
The asymmetry of the spectrum with $H \parallel a$ in Na$_{0.42}$CoO$_{2}\cdot$ 1.3H$_{2}$O is due to the fact that the NQR frequency tensor is not symmetric.  

Figure 2 shows the temperature dependence of the Knight shift ($K$) for both hydrated and unhydrated samples,  
$K_a$ with the field applied along the $a$-axis, and $K_c$ with the field applied along the $c$-axis. 
$K_c$ was determined from the central peak, which agrees well with that determined  from the midpoint between the two first satellites.
$K_a$ was determined from the central peak \cite{note}, by taking into account the shift due to the nuclear quadrupole interaction \cite{Abragam}. For both samples and along  both crystal-axis directions, the Knight shift decreases with decreasing temperature down to $T_0\sim$60 K,
and then becomes a constant at lower temperatures. The most striking feature is that {\it both the magnitude and the temperature dependence of the Knight shifts are identical for the two compounds}, except the drop of $K_{a,c}$ at the superconducting transition in the hydrated sample.
\begin{figure}[h]
\includegraphics[width=7cm]{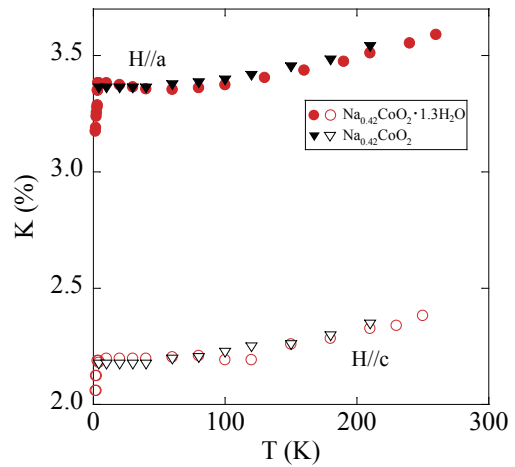}
\caption{\label{fig:shift} (Color online)
Temperature dependence of the Knight shift along $a$- and $c$-axis for hydrated and unhydrated samples. The gyromagnetic ratio of 10.03 MHz/T was used for $^{59}$Co. Data for the hydrated superconductor are from Ref. \cite{Zheng2}.}
\end{figure}

The Knight shift consists of  contributions from the spin susceptibility, $K_s$, and from the orbital susceptibility (Van Vleck susceptibility), $K_{orb}$ \cite{Abragam},
\begin{eqnarray} 
K = K_s + K_{orb} ,
\end{eqnarray}
with $K_{orb}$ being $T$-independent.  $K_s(T)$ and and $K_{orb}$ are respectively related  to the spin susceptibility $\chi_s$ and orbital susceptibility $\chi_{orb}$ as
\begin{eqnarray} 
K_s(T) = A_{hf} \chi_s(T) \\
K_{orb} = A_{orb} \chi_{orb} ,
\end{eqnarray}
where $A_{hf}$ is the hyperfine coupling constant between the nuclear and the electron spins,  and $A_{orb}= 2\langle 1/r^{3}\rangle $ with $\langle ... \rangle$ denoting an average over Co-3d orbit.  
Therefore, the temperature dependence of $K$ is due to the spin susceptibility that decreases with decreasing $T$, which is a common feature seen in low hole-doped \cite{Alloul} and electron-doped \cite{Zheng3} cuprate superconductors. In analogy to the cuprates, this phenomenon may be ascribed to some sort of  pseudogap \cite{Kontani,Li}. However, unlike the cuprates, the seemingly pseudogap is unrelated to superconductivity in the cobaltate.
 
Since $\chi_s$ is proportional to the DOS at the Fermi level, the identical Knight shift in the two compounds therefore indicates that  {\it the DOS is the same in  hydrated and unhydrated samples}. Note that, in the case of Na doping that adds electrons to the Co $t_{2g}$ orbit, DOS is sensitive to the Na content \cite{Zheng}. Therefore, the present result indicates that water intercalation does not alter the carrier concentration, which is at variance with the early suggestion that carrier concentration may change after hydration \cite{Takada2,Milne}. The first-principle calculation has predicted that water intercalation has no effect on band structure other than making the system more two dimensional \cite{Kotliar,Johannes}, which agrees with our finding.
\begin{figure}[h]
\includegraphics[width=7cm]{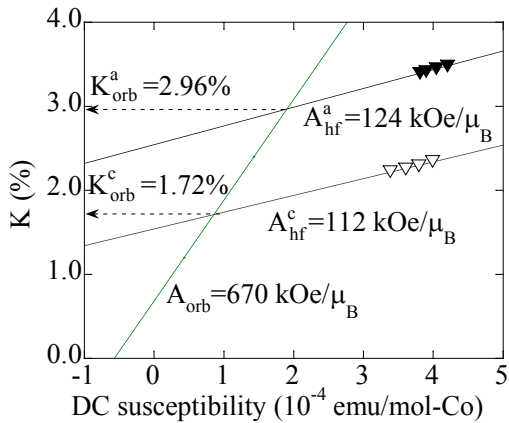}
\caption{\label{fig:t1t} (Color online) $K$-$\chi$ plot for unhydrated Na$_{0.42}$CoO$_{2}$.}
\end{figure}

Next, we estimate the hyperfine coupling constant and separate the spin and orbital susceptibilities.
In Fig. 3, the Knight shift for the unhydrated sample is plotted as a function of DC susceptibility  published previously \cite{Lin}. The diamagnetic susceptibility due to closed shells of Na, Co, and O was estimated to be -5.7$\times$10$^{-5}$ emu$/$mol, from which the slope of $A_{orb}$ was drawn. Here, 
  $\langle 1/r^3 \rangle$=6.7 a.u. is adopted,  which is 80\% theoretical value for Co$^{3+}$ ion \cite{Atomic}.
From this $K-\chi$  plot,  the hyperfine coupling constant  is extracted as 
$A_{hf}^a$=124$\pm$ 20 kOe$/ \mu _B$ and 
$A_{hf}^c$=112$\pm$10 kOe$/ \mu _B$, respectively.
The orbital part of the shift and susceptibility are  
$K^a_{orb}$=2.96 $\pm$ 0.1 \%,
$\chi_{orb}^a$=(2.49$ \pm 0.1)\times$10$^{-4}$ emu $/$ mol$\cdot$Co and
$K^c_{orb}$=1.72 $\pm$ 0.04 \%, 
 $\chi_{orb}^c$=(1.44 $\pm$ 0.03)$\times$10$^{-4}$ emu/mol$\cdot$Co. A previous study using aligned powders, based on a method different from the $K-\chi$ plot,  has given  similar $K^{a,c}_{orb}$; the values obtained there are about 10\% larger than ours \cite{Muk}. 
It is remarkable that, within the experimental uncertainty, {\it the hyperfine coupling constant is nearly isotropic}. For the hydrated sample, the DC susceptibility is contaminated by a Curie upturn, 
which prevents one to do the $K-\chi$ plot. Since the Knight shift is identical to the unhydrated compound, and so is the $1/T_1T$ at $T\geq$60 K (see below),   it is reasonable to expect the same hyperfine coupling constant for the hydrated sample.


\begin{figure}[h]
\includegraphics[width=7cm]{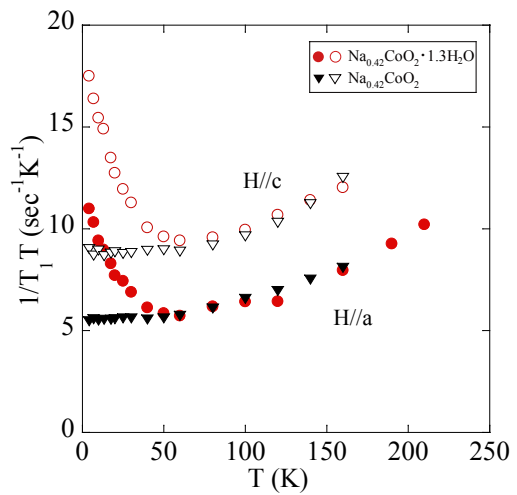}
\caption{\label{fig:kai} (Color online) $T$-dependence of 1/$T_1T$ for hydrated and unhydrated samples. }
\end{figure}

Finally, we evaluate the  spin correlations.
Figure 4 shows the temperature dependence of $1/T_1T$ above 4.2 K for 
 $H \parallel c$ and $H \parallel a$  in both hydrated and unhydrated samples.
Again, it is striking that, {\it above $T_0$=60 K, $1/T_1T$ is identical for both compounds}. The two compounds share the same feature that $1/T_1T$ decreases with decreasing $T$ down to $T_0\sim$60 K in both directions.
Below $T_0$, $1/T_1T$ is constant for unhydrated sample, which indicates that the electronic correlations are  weak. In fact,  the Korringa ratio 
\begin{eqnarray}
S=T_1TK_s^2 \cdot\frac{4\pi k_B}{\hbar}(\frac{\gamma_n}{\gamma_e})^2, 
\end{eqnarray}
which is unity for a free electron system, is 0.65$\pm$0.3, being close to the value for Pb or Be metals.  Therefore, {\it the unhydrated Na$_{0.42}$CoO$_2$ is a conventional metal that well conforms to  Fermi liquid theory}. This is further corroborated by the  good agreement  between  calculated  DOS by LDA (local density-approximation) and our estimated value. The spin susceptibility $\chi_s^a$ is extracted from Fig. 3 to be 1.91$\times$10$^{-4}$ emu $/$ mol$\cdot$Co at 100 K, which yields the DOS at the Fermi level $N(E_F)=\chi_s/2\mu_B^2$=2.95 state/eV. This is very close to the LDA  value of 4.4 state/eV \cite{Singh}. Therefore, the frequently-used  hypothesis that strong correlations are present in the parent compound may need to be re-examined \cite{Zhou,Ishida,Fang}.

In contrast, $1/T_1T$ increases with decreasing $T$ for the hydrated sample in both directions, indicating strong electron correlations.
In a general form, $1/T_1T$  is written as
$\frac{1}{T_1T}= \frac{\pi k_B \gamma^2_n }{(\gamma_e \hbar )^2} \sum_q A_{hf}^2 \frac{\chi ''_{\perp}(q,\omega)}{\omega}$
, where 
$\chi ''_{\perp}(q,\omega)$ is the imaginary part of the dynamical susceptibility  perpendicular to the applied field, and $\omega$ is the NMR frequency.

If one assumes that there is a peak around a finite wave  vector $Q$ (due to spin fluctuation), then one may have the following approximation,
\begin{eqnarray} 
\frac{1}{T_1T}= \left( \frac{1}{T_1T} \right)_0 + \left( \frac{1}{T_1T} \right)_Q \\
\left( \frac{1}{T_1T} \right)_Q = \frac{\pi k_B \gamma^2_n }{(\gamma_e \hbar )^2} \sum_{q \approx Q} A_{hf}^2 \frac{\chi ''_{\perp}(q,\omega)}{\omega},
\end{eqnarray}
where   $\left( 1/T_1T \right)_Q$ is the contribution from  wave vectors around  $Q$, 
 while $\left( 1/T_1T \right)_0$ denotes the contribution from   $q\sim$0. 
The unhydrated sample becomes a conventional metal,  while the spin correlation develops upon decreasing temperature in the hydrated sample.
Therefore, the water intercalation is to bring about the antiferromagnetic-like spin correlations, perhaps through modifying the Fermi surface as to favor nesting conditions. Angle resolve photoemission spectroscopy is encouraged to compare the Fermi surface of hydrated and unhydrated compounds. The insertion of two layers of water molecules largely separates the CoO$_2$ layers, and it is likely because of the hydrogen bonding \cite{Jorgensen,Lynn} that makes water  particularly suitable as a spacer than others. 


We estimate the anisotropy of the spin fluctuations in the superconductor by analyzing $\sum_{q \approx Q} \chi '' (q)$. $\left( 1/T_1T \right)_Q$ is obtained by
 subtracting  $\left( 1/T_1T \right)_0$ for unhydrated sample from the observed $1/T_1T$. From the symmetry of the crystal structure, we have,
\begin{eqnarray} 
\left(\frac{1}{T_1T} \right)_Q^c= 2\frac{\pi k_B \gamma^2_n }{(\gamma_e \hbar )^2 \omega_n} \sum_{q \approx Q} (A_{hf}^c)^2 \chi ''_a \left(q \right) \\
\left(\frac{1}{T_1T} \right)_Q^a= \frac{\pi k_B \gamma^2_n }{(\gamma_e \hbar )^2 \omega_n} \sum_{q \approx Q} \left[ (A_{hf}^a)^2 \chi ''_a \left(q \right)+(A_{hf}^a)^2 \chi ''_c \left(q\right) \right]
\end{eqnarray}
By making use of  nearly isotropic hyperfine coupling constant, the anisotropy of the susceptibility around $Q$ is obtained as shown in Fig. 5.  
As can be seen in the figure, $\sum_{q \approx Q} \chi ''_a (q) \geq 3\sum_{q \approx Q} \chi ''_c (q) $ just above $T_c$. Namely,  the spin fluctuations there in the CoO$_2$ plane are more than three times stronger  than along the $c$-axis direction (inter-plane). Thus, {\it the highly anisotropic spin fluctuations turn out to be the most important ingredient of the superconductor}.

\begin{figure}[h]
\includegraphics[width=8cm]{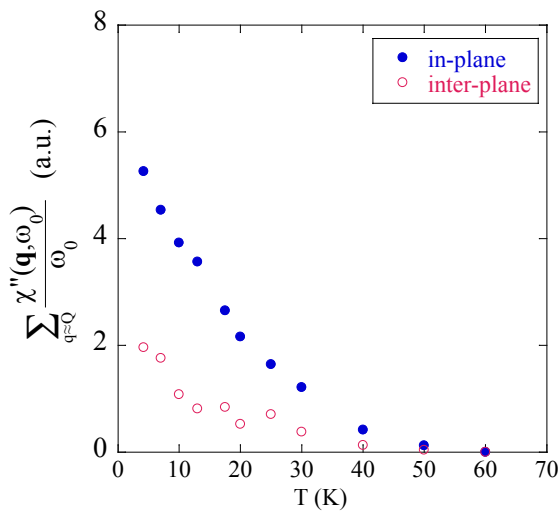}
\caption{\label{fig:kai} (Color online) $T$-dependence of the susceptibility around $Q$. The filled marks represent $\sum_{q \approx Q} \chi ''_a (q)$, while the open marks for $\sum_{q \approx Q} \chi ''_c (q) $.}
\end{figure}

In conclusion, we have presented $^{59}$Co-NMR 
measurements and analysis on the hydrated and unhydrated cobalt oxide  single crystals. The unhydrated compound  Na$_{0.42}$CoO$_{2}$ is found to be a non-correlated metal. 
After water intercalation, the antiferromagnetic-like spin fluctuations develop but there is no change in the DOS at the Fermi level. Therefore, the primary role of  water intercalation is to make the system more two dimensional and brings about spin fluctuations. 
The spin correlations in the superconductor is anisotropic; they are stronger   in the CoO$_2$ plane  than inter-plane by a factor of more than 3, which is quite different from the cuprate case  where isotropic spin fluctuations have been observed. These new results and insights provide a foundation for understanding the  mechanism of superconductivity in cobalt oxides.

This work was supported  in  part  by a research grant from  MEXT, No. 17072005.


\begin{thebibliography}{}
\bibitem{Takada}
K. Takada {\it et al}, Nature {\bf 422}, 53 (2003).

\bibitem{Fujimoto} 
T. Fujimoto, G. - q. Zheng, Y. Kitaoka, R.L. Meng, J. Cmaidalka and C.W. Chu, Phys. Rev. Lett. {\bf 92}, 047004 (2004).

\bibitem{Zheng}
G. - q. Zheng, K. Matano, R.L. Meng, J. Cmaidalka, and C.W. Chu, J. Phys.: Condens. Matter {\bf 18}, L63 (2006).

\bibitem{Kusano}
E. Kusano {\it et al}, Phys. Rev.  B {\bf 76}, 100506 (R) (2007).


\bibitem{Zheng2}
G. - q. Zheng, K. Matano, D. P. Chen, and C. T. Lin,
Phys. Rev. {\bf B 73},180503 (R) (2006).

\bibitem{Kobayashi}
Y. Kobayashi, {\it et al}, J. Phys. Soc. Jpn.  {\bf 74},  1800 (2005).

\bibitem{Eremin}
M.M. Korshunov and I. Eremin, arXiv:0708.0807.

\bibitem{ZhouWang}
 Sen Zhou and Ziqiang Wang, arXiv:0712.1042.


\bibitem{Kastner}
M. A. Kastner{\it et al}, Rev. Mod. Phys.  {\bf 70}.  897 (1998).


\bibitem{Kotliar}
C.A. Marianetti, G. Kotliar and G. Ceder, Phys. Rev. Lett. {\bf 92}, 196405 (2004).

\bibitem{Johannes}
M.D. Johannes and D.J. Singh, Phys. Rev. B {\bf 70}, 014507 (2004).

\bibitem{Takada2}
K. Takada {\it et al.}, J. Mater. Chem. {\bf 14}, 1448 (2004).

\bibitem{Milne}
C. J. Milne {\it et al}, Phys. Rev. Lett. {\bf 93}, 247007 (2004).

\bibitem{Muk}
I. R. Mukhamedshin {\it et al}, Phys. Rev. Lett. {\bf 94},  247602 (2005).

\bibitem{Ning}
F.L. Ning {\it et al}, Phys. Rev. Lett. {\bf 93}, 237201 (2004).

\bibitem{IshidaK}
K. Ishida, Y. Ihara, H. Takeya, C. Michioka, K. Yoshimura, K. Takada, T. Sasaki, H. Sakurai, E. Takayama-Muromachi, Physica C {\bf 460-462}, 190 (2007). 


\bibitem{Chen}
D.P. Chen {\it et al},
Phys. Rev.{\bf B 70}, 0245506 (2004).

\bibitem{Lin}
C.T. Lin {\it et al},
Physica C: {\bf 460-462} 471 (2007).


\bibitem{Narath}
A. Narath ,Phys. Rev. {\bf162}, 320 (1967).

\bibitem{note}
Here we take the opportunity to correct an error in calculating the effect of nuclear quadrupole interaction which appeared in the  data of $K_a$ ($T\geq$ 100 K) of Ref.\cite{Zheng2}. 

\bibitem{Abragam}
A. Abragam, {\it Principles of Nuclear Magnetism}  (Oxford University Press, New York, 1961).

\bibitem{Alloul}

 M. Takigawa {\it et al}, 
 Phys. Rev. B {\bf 43}, 247 (1991)


\bibitem{Zheng3}
G.-q. Zheng {\it et al}, Phys. Rev. Lett. {\bf 90}, 197005 (2003).

\bibitem{Kontani}
K. Yada and H. Kontani, J. Phys. Soc. Jpn. {\bf 74}, 2161 (2005).

\bibitem{Li}
Z.-J. Yao,  J.-X. Li,  and Z. D. Wang, Phys. Rev. B {\bf 76}, 212506 (2007).



\bibitem{Atomic}
S. Fraga {\it et al}, {\it
Handbook of Atomic Data} (Elsevier Pub., Amsterdam, 1976).



\bibitem{Singh}
D.J. Singh, Phys. Rev. B {\bf 61}, 1339 (2000).

\bibitem{Zhou}
S. Zhou {\it et al}, Phys. Rev. Lett. {\bf 94}, 206401 (2005).

\bibitem{Ishida}
H. Ishida, M.D. Johannes, A. Liebsch, Phys. Rev. Lett. {\bf 94}, 196401 (2005).

\bibitem{Fang}
G.-T. Wang, X. Dai and Z. Fang, arXiv:0801.41841.

\bibitem{Jorgensen}
J. D. Jorgensen {\it et al.}, Phys. Rev. {\bf B 68}, 214517 (2003).

\bibitem{Lynn}
J. W. Lynn {\it et al.}, Phys. Rev. {\bf B 68}, 214516 (2003).


\end{thebibliography}

\end{document}